\documentclass[iop]{emulateapj}

\providecommand{\mb}{$\Delta m_{15}(B)$}

\providecommand{\mu}{$\Delta m_{15}(U)$}

\begin{document}
\title{Swift Ultraviolet Observations of Supernova 2014J in M82: \\
	Large Extinction from Interstellar Dust}
\author{Peter~J.~Brown\altaffilmark{1}, Michael T. Smitka\altaffilmark{1}, Lifan Wang\altaffilmark{1,2}, Alice Breeveld\altaffilmark{3}, \\  Massimiliano de Pasquale\altaffilmark{4},
Dieter H. Hartmann\altaffilmark{5}, 
Kevin Krisciunas\altaffilmark{1}, N. P. M. Kuin\altaffilmark{3},\\ Peter A. Milne\altaffilmark{6}, Mat Page\altaffilmark{3}, \& Michael Siegel\altaffilmark{7} }
\altaffiltext{1}{George P. and Cynthia Woods Mitchell Institute for Fundamental Physics \& Astronomy,
Texas A. \& M. University, Department of Physics and Astronomy, 
4242 TAMU, College Station, TX 77843, USA }            
\altaffiltext{2}{Purple Mountain Observatory, Chinese Academy of Sciences, Nanjing 210008, China }
\altaffiltext{3}{Mullard Space Science Laboratory, University College London, Holmbury St. Mary, Dorking Surrey, RH5 6NT, UK}            
\altaffiltext{4}{Instituto di Astrofisica Spaziale e Fisica Cosmica di Palermo
Via Ugo la Malfa 153
90146 Palermo
Italy}
\altaffiltext{5}{Clemson University, 
Department of Physics \& Astronomy, 
Kinard Lab of Physics, 
Clemson, SC 29634-0978, USA}
\altaffiltext{6}{Steward Observatory, University of Arizona, Tucson, AZ 85719, USA}            
\altaffiltext{7}{Department of Astronomy and Astrophysics, 
				The Pennsylvania State University, 
				525 Davey Laboratory, 
				University Park, PA 16802, USA}

\begin{abstract}

We present optical and ultraviolet (UV) photometry and spectra of the very nearby and highly reddened supernova (SN) 2014J in M82 obtained with the Swift Ultra-Violet/Optical Telescope (UVOT).  Comparison of the UVOT grism spectra of SN~2014J with Hubble Space Telescope observations of SN2011fe or UVOT grism spectra of SN~2012fr are consistent with an extinction law with a low value of R$_V\sim$1.4.  The high reddening causes the detected photon distribution in the broadband UV filters to have a much longer effective wavelength than for an unreddened SN.   The light curve evolution is consistent with this shift and does not show a flattening due to photons being scattered back into the line of sight.  The light curve shapes and color evolution are inconsistent with a contribution scattered into the line of sight by circumstellar dust.  We conclude that most or all of the high reddening must come from interstellar dust.
We show that even for a single dust composition, there is not a unique reddening law caused by circumstellar scattering.  Rather, when considering scattering from a time-variable source, we confirm earlier studies that the reddening law is a function of the dust geometry, column density, and epoch.  We also show how an assumed geometry of dust as a foreground sheet in mixed stellar/dust systems will lead to a higher inferred R$_V$.  Rather than assuming the dust around SNe is peculiar, SNe may be useful probes of the interstellar reddening laws in other galaxies.

\end{abstract}


\section{Introduction \label{intro}}

Type Ia Supernovae (SNe) are important cosmological tools because their optical/near-infrared luminosities are intrinsically bright and predictable.  This means that their intrinsic brightness is correlated with distance-independent colors and light curve shapes (Phillips 1993, Riess et al. 1996, Goldhaber et al. 2001) so that they can be used as ``standardizable'' candles.  
Equally important (though perhaps implicit in being standard candles in multiple filters) is the fact that their colors are also predictable.  

Understanding the colors allows one to infer the amount of dust extinction affecting the brightness in each of the filters based on the amount it preferentially extinguishes emission at shorter wavelengths, i.e. the amount of reddening.  
The amount of extinction (in magnitudes) at a certain wavelength or filter is often expressed as A$_{\lambda}$~=~R$_{\lambda}~\times$~E(B-V).
R$_{\lambda}$ is the extinction coefficient for that wavelength or filter.
E(B-V) is the differential extinction between the B and V filters, also called the color excess:  
E(B-V)= A$_B$-A$_V$=(B-V)$_{observed}$-(B-V)$_{intrinsic}$.  
E(B-V) is often used to parameterize the amount of dust, though the observed value also depends on both the source spectrum and the shape of the extinction law \citep{McCall_2004}.  
The equation A$_V$~=~R$_V~\times$~E(B-V)
shows the relationship between the B-V reddening and the extinction in V.  Reddening in the Milky Way (MW) has an average value of R$_V\sim3.1$ \citep{Weingartner_Draine_2001}.  A different extinction law could have a higher R$_V$ value, resulting in more extinction for the same amount of B-V reddening.  By definition, R$_B$ = R$_V$ + 1, so extinction laws with higher R$_V$ are sometimes called shallower because of the smaller relative difference between A$_B$ and A$_V$.  Conversely, extinction laws with lower R$_V$ have a relatively steeper change in the extinction with wavelength.

Curiously, a low value value for R$_V$ has been found for highly reddened individual SNe Ia, including SNe 1999cl: 2.01 \citep{Krisciunas_etal_2006};  2003cg: 1.80 \citep{Elias-Rosa_etal_2006}; 2006X: 1.48 \citep{Wang_etal_2008}, and large samples of SNe Ia, including \citet{Tripp_1998}: 2, \citet{Reindl_etal_2005}: 2.65,\citet{WangX_etal_2006}: 2.3, \citet{Conley_etal_2007}: 1, \citet{Kessler_etal_2009}: 2.18, \citet{Hicken_etal_2009_DE}: 1.7.
The origin and implications of this unusual extinction law are uncertain. 
It could be due to small dust grains; such low R$_V$ are rare but have been observed in the MW \citep{Gordon_etal_2003}.  \citep{Cardelli_etal_1989} parameterized a family of extinction curves as a function of R$_V$, but the smallest value of R$_V$ used was 2.6.  So the use of smaller values is an extrapolation of the relationship between the reddening law shape and the R$_V$.
 \citet{Wang_etal_2009_HV} find that when separated based on their expansion velocities, the absolute magnitudes of high-velocity (HV) SNe Ia tend to follow the steeper extinction law, while the others are more consistent with MW (i.e. R$_V$=3.1) dust.  \citet{Foley_etal_2011_hv} argue that it is only the highly reddened SNe Ia which follow the steep dust law, while HV SNe Ia have different intrinsic colors, and the less reddened samples of both populations can be fit with an extinction law similar to that of the MW.

\citet{Wang_2005} and \citet{Goobar_2008} show that circumstellar scattering will modify the ``intrinsic'' extinction law, resulting in a lower value of R$_V$.  The extinction laws shown in \citet{Goobar_2008} were derived for a constant luminosity source.  The effective extinction law for a time-varying SN would be more complicated and depend on the dust geometry and density as well as the intrinsic emission from the SN \citep{Wang_2005,Amanullah_Goobar_2011}.  But the scattering of light could result in a smaller value of R$_V$.

Ultraviolet emission is very sensitive to the dust grain size and geometry.  While the amount of extinction in the mid-UV is less dependent on the grain size when normalized by the color excess E(B-V) \citep{Cardelli_etal_1989}, the relative extinction compared to the optical is very dependent on the grain size.  
For the circumstellar scattering case, the wavelength dependence of scattering and absorption steepens the effect of extinction in the UV \citep{Wang_2005,Goobar_2008,Brown_etal_2010}.  This is because the albedo for MW/LMC type grains peaks in the optical, reducing the extinction in the optical while increasing E(B-V) for the same amount of dust. Observations on either side of the albedo bump should reveal whether the extinction law shapes are due to dust variations or circumstellar scattering.  

The high sensitivity of UV photons to extinction makes them a good probe of the extinction laws, but it also makes them very hard to detect.  Once there is enough dust (of whatever form or geometry) to make a significant impact on the optical colors, the UV light is almost completely extinguished. It is a challenge to detect extinguished UV emission at the distances at which SNe are typically discovered.
The discovery of the very nearby, and highly reddened, SN~2014J in M82 \citep{Fossey_etal_2014} provides a unique opportunity to study the effect of heavy extinction on the UV light.  However, we will show that measuring these effects is difficult in the UV, especially with broad-band filters.  

In Section \ref{obs} we present UV/optical photometry and spectroscopy of SN~2014J obtained with the Ultra-violet/Optical Telescope (UVOT; \citealp{Roming_etal_2005}) on the Swift satellite \citep{Gehrels_etal_2004}.    We compare different extinction laws to the observed spectra in Section \ref{extinctionlaws}.   Section \ref{model} describes our circumstellar model and Section \ref{comparison} compares it to the observed photometry.  In Section \ref{galaxyextinction} we show how the assumption of dust extinction as a foreground screen increases the inferred R$_V$.  We summarize these results in Section \ref{discussion}.

 
\section{Swift Observations of SN~2014J in M82 }\label{obs} 

The Swift satellite started observing SN~2014J in M82 \citep{Cao_etal_2014, Fossey_etal_2014} at 2014-01-22.43 UT. Early Swift results were reported by \citet{Brown_Evans_2014}.
Photometry observations were made on a roughly daily basis between Jan 22 and Feb 25 and then continuing with a lower cadence until April 10.  Because the UVOT works as a photon detector, bright sources suffer from coincidence loss--multiple photons arriving during the same frame are undercounted (see \citealp{Poole_etal_2008} for more details and the correction method).  Because of the brightness of the SN in the optical, special ``hardware window'' modes were used to more frequently read out a smaller section of the detector to reduce the effect of coincidence loss due to the bright SN and the relatively bright underlying galaxy.  Because of the faintness in the UV additional observations were made to increase the exposure time in the uvm2 filter.  The filter characteristics are described in \citet{Poole_etal_2008} and \citet{Breeveld_etal_2011}, but will be described in detail below.
The host galaxy was imaged repeatedly by UVOT in the years prior to explosion \citep{Hutton_etal_2014}.  
These images were used to subtract the count rates of the underlying galaxy.

Observations with the UVOT's UV grism were made at seven epochs before and around maximum light (though each epoch is broken up into separate exposures due to the observability windows).  The time ranges and total exposure times used here are listed in Table \ref{table_grism}.  
We also use the Swift/UVOT grism spectra of SN~2012fr for comparison purposes  (PI: R. Foley).  SN~2012fr was a very broad NUV-red/irregular SN with high velocity features. We use a distance modulus of 31.18 $\pm$  0.05 derived from Cepheid observations \citep{Freedman_etal_2001}.  The observations used are also listed in table \ref{table_grism}. 
\begin{figure} 
\includegraphics[scale=1]{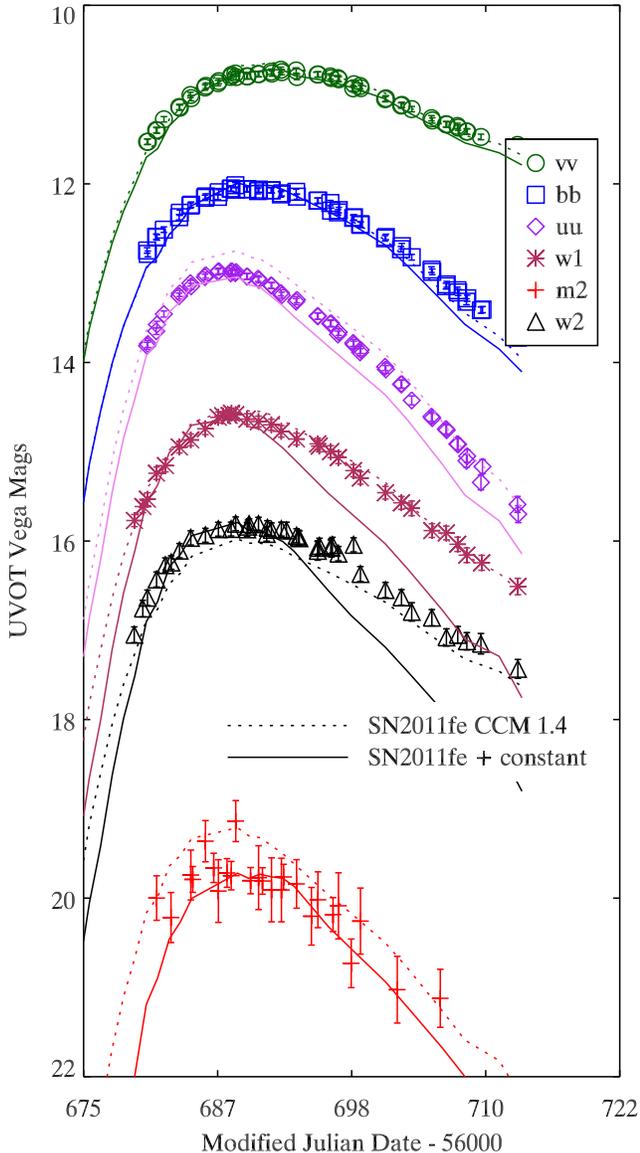} 
\caption[Results]
        {UVOT light curves of SN2014J given in observed Vega magnitudes versus time.  The reddening naturally separates the curves with no arbitrary offsets needed.  We overplot the spectrophotometric light curves for SN~2011fe (based on bolometric spectra from \citealp{Pereira_etal_2013}) without extinction (but offset to match the peak magnitudes).  We also plot spectrophotometric light curves for SN~2011fe after reddening with a CCM extinction law with E(B-V)=1.4 and R$_V$=1.4 
and correcting only for the difference in distance moduli.
 }\label{fig_lightcurves}    
\end{figure}

\subsection{Photometric Data Reduction}
Swift UVOT data were analyzed using the reductions methods for the Swift Optical/Ultraviolet Supernova Archive (SOUSA; \citealp{Brown_etal_2014_SOUSA}) including the revised UV zeropoints and time-dependent sensitivity from \citet{Breeveld_etal_2011}. 
 The photometry is given in Table \ref{table_photometry}.  The six filter light curves  are displayed in Figure \ref{fig_lightcurves}.  

 Because of the brightness of the SN and the underlying galaxy, we do not use the normal full-frame images in the optical but only those taken using a hardware window to increase the frame rate (by reading out a smaller section of the detector) and reduce coincidence losses.  
By comparison with well calibrated SN photometry from ground based observations with linear detectors, \citet{Brown_etal_2009} found the coincidence loss correction to yield accurate magnitudes in the full field mode when the underlying galaxy count rate is less than about 5 counts s$^{-1}$ in a 5\arcsec aperture. 
 The underlying galaxy count rates in b and v are more than five times the brightness limit for the full-frame images, so the factor of four gained by using the smaller readout windows is not enough in this case.
The optical bv data is found to be about 0.2 mag fainter at peak than that found by \citet{Tsvetkov_etal_2014} and \citet{Marion_etal_2014}. 
 The count rates of the host galaxy in the other UVOT filters are much lower than the optical and thus the photometry should not be significantly affected.  In the rest of the paper we use the B and V photometry from \citet{Marion_etal_2014} which is similar to the UVOT bv photometry of other SNe against which we have compared \citep{Poole_etal_2008,Brown_etal_2009}.  
In many instances we reference the time of observations with respect to the time at which the SN reached maximum light in the B band (t$_{Bmax}$).  We use the SNooPy fit results from \citet{Marion_etal_2014} giving the time of maximum light as MJD 56689.74 (2014 Feb 1.74).
For SN~2014J we use a distance modulus of 27.73 $\pm$ 0.02  derived from color-magnitude fitting of the tip of the red giant branch \citep{Jacobs_etal_2009}.

To measure the light curve parameters in the UV filters, the light curves have been fit by stretching appropriate templates to the observed data.  The three UV filters are fit with the UVOT templates from SN~2011fe \citep{Brown_etal_2012_11fe} and the u band is also fit using the uvw1 template.  The resulting peak times, peak magnitudes, and the change of magnitude in the 15 days after maximum light are reported in Table  \ref{table_fits}.  The b and v are not reported due to the possible coincidence loss issues. 
\citet{Milne_etal_2010} studied the UV light curve shapes of a large number of Swift SNe Ia.  The UV light curves of SN~2014J are broader than all those in \citet{Milne_etal_2010} with the u band comparable only to SN~2005cf.  SN~2011aa is comparably broad \citep{Brown_etal_2014}.
As shown in Figure \ref{fig_lightcurves}, this broadening is consistent with the extinction causing a photon distribution biased to the redder photons (which fade slower) within each band.  This is discussed in more detail in the Appendix. 


\begin{figure*} 
\plottwo{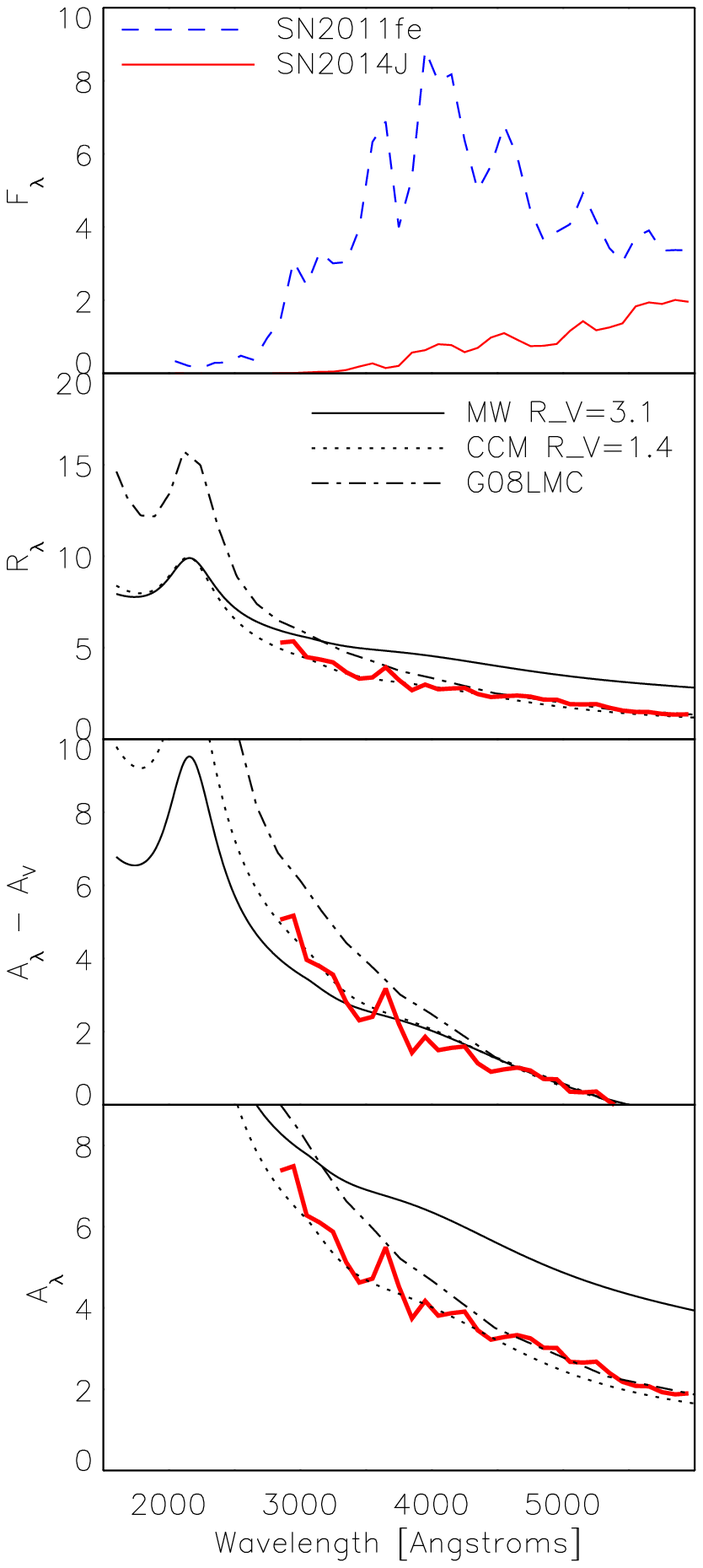}{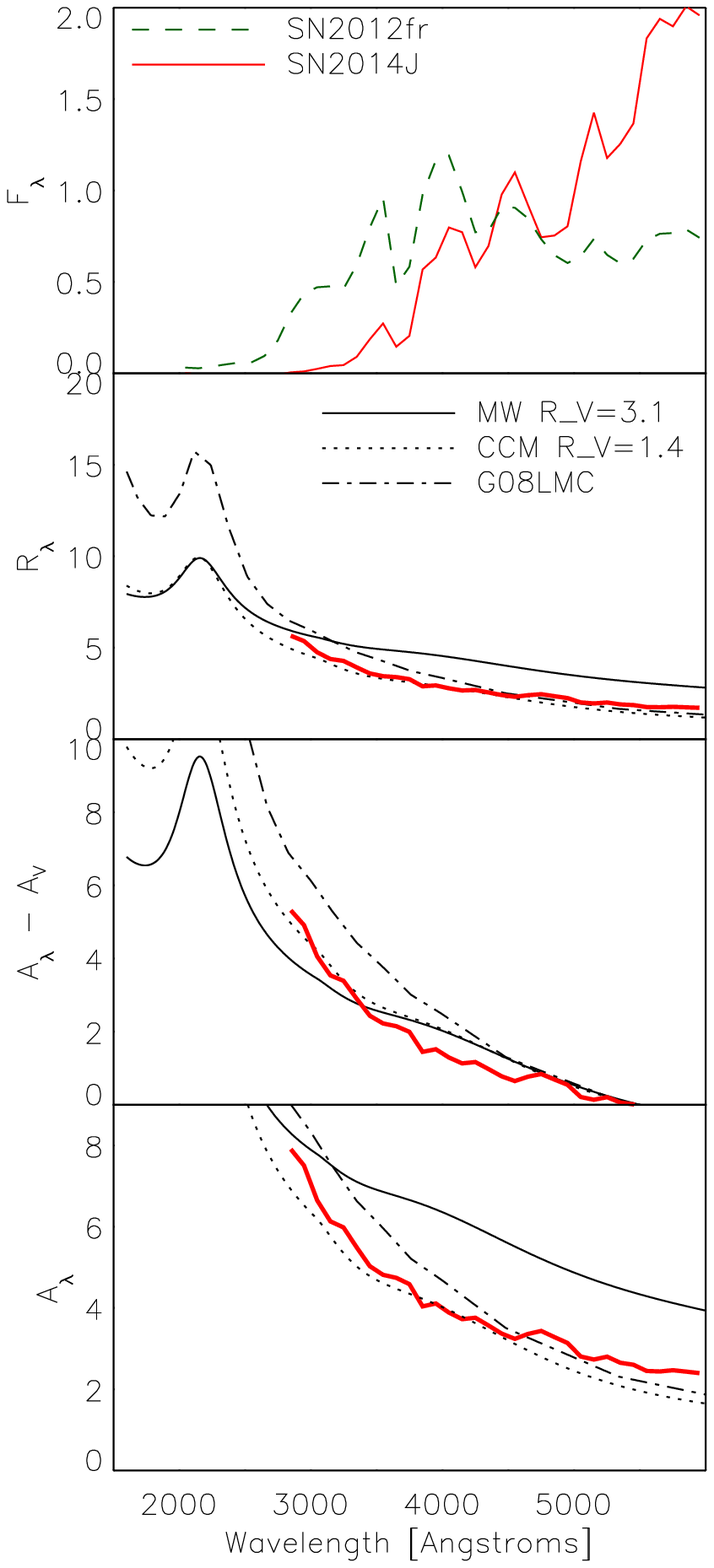}
\caption[Results]
        {Top Left Panel: : Combined spectra of SN~2014J and 2011fe smoothed to 100 A.  
	Lower Left Panels: The spectrum of SN~2014J divided by the spectrum of SN~2011fe and corrected for the relative distances to mimic R$_\lambda$ (i.e. A$_\lambda$ / E(B-V) ), A$_\lambda$-A$_V$, and A$_\lambda$.  These are compared to the corresponding functions for  various extinction laws.
The best match is with the Milky Way law with R$_V$=1.4. 
Right Panels: Same as left but comparing to SN~2012fr.
} \label{fig_specex}    
\end{figure*}

\subsection{Grism Spectral Reduction}

Grism observations of SN~2014J and SN~2012fr  were extracted using the default parameters of the UVOTPY package (Kuin et al. 2014, in preparation).\footnotetext{www.mssl.ucl.ac.uk/www\_astro/uvot} The nominal wavelength accuracy is 20 \AA~ and the flux calibration is accurate to about 10\%.  Second order contamination is small for unreddened, UV-faint SNe Ia and considered negligble in this case.  The exposures from each day (corresponding to the ExposureID's in Table \ref{table_grism}) were wavelength shifted to agree with a master frame using a chi square minimizing wavelength shift routine.  Next, they were coadded using variance weighting to create a single spectrum for each obsid.  The SN~2012fr spectra were then wavelength shifted  to agree with a maximum light spectrum of 2012fr from \citet{Childress_etal_2013} obtained from the WISEREP database \citep{Yaron_Gal-Yam_2012} between wavelengths of 3500-5500 \AA. 
For comparison of the spectra near maximum-light, SN~2014J spectra between Feb 1-5 were combined into a single spectrum as were both SN~2012fr spectra.
The combined maximum light spectra are displayed in the top panel of Figure \ref{fig_specex}.

\section{Comparing Extinction Laws to the Maximum Light Spectra}\label{extinctionlaws}

The extinction law, by which we mean the wavelength dependence of extinction, has been measured to many lines of sight in the Milky Way (MW), the Large Magellanic Cloud (LMC) and the Small Magellanic Cloud (SMC).  Cardelli et al. (1989; CCM) parameterized the different shapes of the extinction law by the corresponding R$_V$.  In this work we will consider extinction laws with R$_V$=3.1 corresponding to the average value of the MW, as well as other values of R$_V$, in particular R$_V$=1.4, the best fit value from \citet{Amanullah_etal_2014}.  These will be called ``MW 3.1'' and ``CCM 1.4'' hereafter.  They correspond to different slopes in the optical and near-UV.  Both feature the ``2175 \AA'' bump, whose strength (as measured by R$_{2175 \AA}$) is largely independent of R$_V$ \citep{Cardelli_etal_1989}. 
We also compare with an LMC extinction law modified by the affects of circumstellar scattering (for a steady state source, dubbed G08LMC) from \citet{Goobar_2008}.  Circumstellar scattering will be discussed in more detail below.  

To infer the wavelength dependent extinction, one needs a comparison object with similar intrinsic colors.  To first order, SNe Ia appear to follow a single-parameter continuum determined by the $^{56}$Ni yield.  The observed parameter typically associated with this is the light curve width or the decay parameter \mb, which is a measurement of the magnitudes the B-band light curve decays in the 15 days after maximum light.  One would typically compare SNe Ia with similar values of \mb.  This is somewhat complicated by the extinction because the evolving spectral shape results in a different decay rate when extinguished.  \citet{Phillips_etal_1999} give this correction  as:  \mb$_{true}$=\mb$_{obs}$+0.1 $\times$ E(B-V).             
\citet{Tsvetkov_etal_2014} report a \mb=1.01 and E(B-V)=1.3 for SN~2014J, resulting in an estimated reddening corrected \mb=1.14.  

SN~2011fe is a SN with \mb=1.108 \citep{Munari_etal_2013} with excellent multi-wavelength observations, e.g. from \citet{Brown_etal_2012_11fe,Pereira_etal_2013,Hsiao_etal_2013,Mazzali_etal_2014}.
\citet{Goobar_etal_2014}, \citet{Amanullah_etal_2014}, and \citet{Foley_etal_2014} found it to be a good comparison for SN~2014J.  
We also compare it to SN~2012fr, which is a broad SN Ia  \mb=0.8 \citep{Zhang_etal_2014} with high-velocity spectral features \citep{Childress_etal_2013}.

To determine the observed reddening law we use a maximum light HST spectrum for SN~2011fe \citet{Mazzali_etal_2014}, a combined maximum light UVOT spectrum for SN~2012fr, and the combined maximum light UVOT spectrum for SN~2014J.  We smooth the spectra by rebinning them to 100 \AA~resolution while conserving the flux.  The smoothed spectra are shown in the top panels of Figure \ref{fig_specex}, with SN~2011fe on the left and SN~2012fr on the right.  To determine the extinction difference as a function of wavelength we correct for distance and divide the flux of the reference spectrum by the flux of SN~2014J and convert it to an extinction in magnitudes.  The wavelength dependence of extinction can be visualized in different ways, so in the lower panels we show R$_\lambda$ (i.e. normalizing by E(B-V) to focus on the shape), A$_\lambda$-A$_V$ (to focus on the slope differences), and  A$_\lambda$ (the total effect).
In all panels we show the respective values for the various extinction laws using E(B-V)=1.4, the approximate color excess of SN~2014J from \citet{Amanullah_etal_2014}.  One could also correct to A$_V$ or any other parameter.  

The observed extinction laws, as determined by both comparison spectra, are very similar.  The one derived from SN~2011fe shows more structure due to the large velocities in SN~2014J.  In all plots, but best seen in terms of A$_\lambda$ in the bottom panels, the CCM 1.4 law gives the best match.  The G08LMC law is the next best fit, though with the measured extinction trending lower than it shortward of 4000 \AA.  Better fits might be obtainable with the CCM $R_V$ \citep{Cardelli_etal_1989} and power-law \citep{Goobar_2008} parameterizations, but here we were trying to test the fit of previously used extinction laws.  
Deviations could result from the difference between the real and model extinction curves or the true, unreddened SN and the SN model.

\begin{figure} 
\includegraphics[scale=0.8]{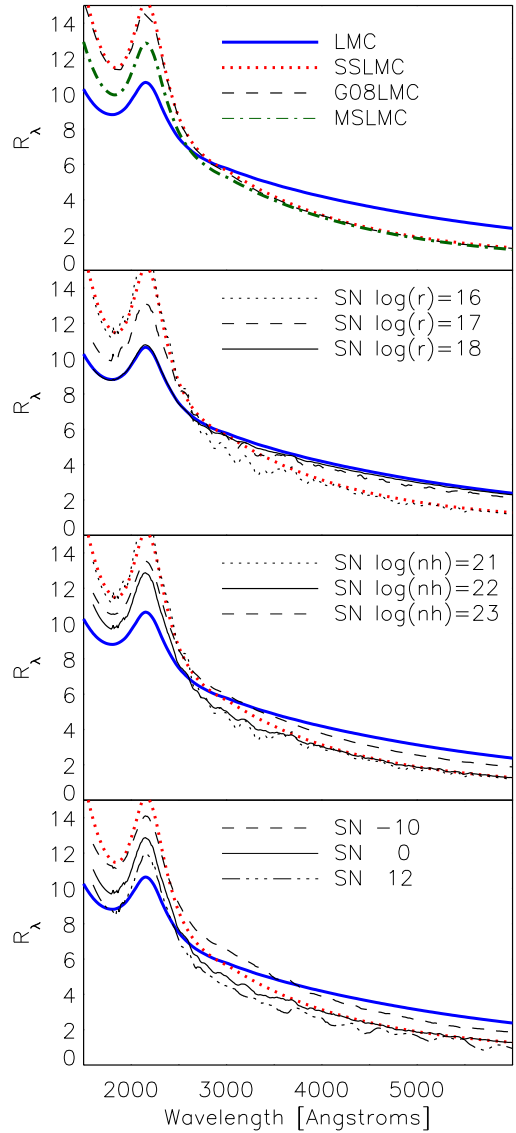} 
\caption[Results]
        {Top Panel: Extinction law for the LMC compared to a single scattering model (SSLMC), the G08 model, and a multiple scattering model (MSLMC).  
Lower Panels:  the effective extinction laws are shown for different radii (in log(cm) at maximum light with a constant optical depth), hydrogen column density (at maximum light with a constant radius), and epochs (in days from maximum light, with a constant radius and optical depth).  Clearly a single law cannot be used for circumstellar scattering,even for identical dust properties.  A grid of laws for various geometries, densities and epochs is needed.  
 }\label{fig_csmultiplotm}    
\end{figure}

\section{Circumstellar Scattering Model} \label{model}

Many lines of evidence show that the reddening law to SN~2014J has a low value of R$_V$.  These include the photometric colors shown above from Swift/UVOT as well as those of HST \citep{Amanullah_etal_2014}.  The near-UV/optical spectroscopy in Section \ref{extinctionlaws} and \citet{Foley_etal_2014} and optical spectroscopy from \citep{Goobar_etal_2014} also show a low R$_V$.   
A low value of R$_V$ could be caused by smaller dust grains \citep{Weingartner_Draine_2001} or from certain geometries of circumstellar dust \citep{Wang_2005,Goobar_2008}.  One key difference is in the temporal changes in the extinction magnitudes and colors.  In the case of interstellar scattering from dust grains, the only temporal changes are due to the changing spectral shape.  R$_\lambda$ will change modestly as the spectral shape changes \citep{Phillips_etal_1999,Wang_etal_2005}.  In the case of circumstellar scattering, however, photons are scattered back into the line of sight with a delay time 
proportional to the distance to the scattering dust \citep{Wang_2005, Amanullah_Goobar_2011}.  As SN~2014J is shown to have a very low value of R$_V$ (\citealp{Amanullah_etal_2014,Foley_etal_2014} and above), we now test the other predictions of scattering to see if they are also consistent.

For extinction from a foreground cloud or sheet of dust, scattering and absorption both remove photons from the line of sight, and the flux is scaled by the factor $e^{-\tau}$, where tau is the optical depth due to scattering and absorption.   For a cloud around the source, just as much light is scattered back into the line of sight as is scattered out of it.  
The fraction of the optical due to scattering is given by the albedo $\omega$.  In the single scattering approximation (each photon can be absorbed or scattered at most once) this scattered flux is added back in and only the absorbed photons are lost.  The observed flux (F$_{obs}$) is thus related to the intrinsic flux (F$_{int}$) as follows:
\begin{equation}
F_{obs}=F_{int}\times(e^{-\tau} + \omega\times(1- e^{-\tau}))
\end{equation} 
In the analysis below we use the wavelength-dependent values\footnote{https://www.astro.princeton.edu/~draine/dust/dustmix.html} of $\tau$ and $\omega$ consistent with average properties of LMC dust \citet{Weingartner_Draine_2001}.  We will refer to this single scattering extinction law as SSLMC below. For low values of $\tau$ this is consistent with the analytic expression used by \citet{Wang_2005} and the Monte-Carlo determined law from \citet{Goobar_2008} for the same LMC type dust.   We point out that the power law parameterization of \citet{Goobar_2008} is not directly predicted from circumstellar scattering but is a good fit to the region longward of the 2175 \AA~bump. Because the albedo peaks in the optical, the extinction is reduced much more in the optical than in the UV.

At higher values of optical depth the single-scattering approximation breaks down.  To account for multiple scattering in an analytic way, we use equation 9 of \citet{Mathis_1972} for the fractional flux emitted by the circumstellar shell around an embedded star.  This is given as
\begin{equation}
F_s=1.1[1-e^{(-\tau)}]\times\omega\times~e^{[-\tau(1-\omega)^{0.5}\times(1-g)^{0.3}]}.
\end{equation} 
Here  g is  the scattering phase function, which determines the relative fractions of photons scattered into different angles.
The total flux observed is then \begin{equation}
F_{obs}=F_{int}*[e^{-\tau}+ F_s]
\end{equation} 
 where the first term is the extinguished light directly from the source and the second is the contribution of scattering from the nebula.  This is labeled MSLMC below.  For low values of optical depth this curve matches the shape of the W05 and G08 laws above 3000 \AA.  These modifications of the LMC law are also compared in the top panel of Figure \ref{fig_csmultiplotm}.  In our implementation, the column density of hydrogen (nh) is given as an input to calculate $\tau$ from the cross sections given by \citet{Weingartner_Draine_2001}.

Several groups have used the power-law parameterization of \citet{Goobar_2008} and found it consistent with the observed optical and NIR colors of highly-reddened SNe.  In such instances the \citet{Goobar_2008} parameterization is used almost exclusively as a reddening law -- namely a wavelength dependence of the extinction. 
 However, those scattering laws shown in the top panel of Figure \ref{fig_csmultiplotm} are applicable for circumstellar scattering source with a constant luminosity/spectrum.  As such they cannot be applied to a time-variable source and be considered as a scattering model.  The delay time dependence of the scattering is a critical element of such a model with strong observational consequences \citep{Wang_2005,Amanullah_Goobar_2011}. 

To evaluate the temporal evolution of the scattering, we generalize the model used in \citet{Wang_2005}.  We use the \citet{Pereira_etal_2013} UV/optical spectral series of SN~2011fe interpolated to a 10 \AA~resolution so that that accurate spectrophotometry can be done on the reddened/scattered spectra taking the full filter bandpasses into account. 
Given a dust shell radius (in cm) and optical depth (calculated from the number of hydrogen atoms along the path, n$_h$), the light travel time to the dust shell and back into the line of sight is calculated in 500 radial directions between 0 and 180 degrees.  For each time step in the spectral series, the spectrum at the appropriate time in the past is interpolated.  The flux from the shell escaping into the line of sight is calculated using equation 9 of \citet{Mathis_1972}, taking into account the scattering angle, optical depth, and albedo.  For each direction bin the \citet{Henyey_Greenstein_1941} function is used to determine the fraction of scattered photons which are redirected to the observer.   These time-delayed contributions are integrated over a 360 degree rotation to cover the whole sphere.  While the largest time delay is twice the light travel time to the dust sphere (photons traveling directly away from the observer and being scattered back), the largest solid angle has half that time.  Photons are not scattered isotropically, and the forward-scattering nature of the LMC dust decreases the median delay time.  The contribution from scattered photons is added to the extinguished spectrum for that epoch. This creates a model spectral series representing the observable spectra from a SN like SN~2011fe and the input dust density and geometry.

By comparing the input and output spectral series, we can determine the effective reddening law at any epoch.  The middle two panels of Figure \ref{fig_csmultiplotm} show the reddening laws at maximum light for the SN2011fe template for different radii and column densities.   In \citet{Goobar_2008} the extinction law was considered insensitive to the size of the scattering cloud (with a change in density allowing for the same optical depth).  For a time-variable source, however, the distance to the cloud directly affects the time delay in the scattered photons which will contribute to the observed light curve. The shape of our extinction curve also changes with column density, while \citet{Goobar_2008} showed R$_V$ to be constant with E(B-V).  We do not know whether this difference is due to our multiple scattering approximation or our inclusion of the time variable source.
The bottom panel of Figure \ref{fig_csmultiplotm} shows the effect of this scattering on the effective extinction law at different epochs.  Here the extinction curves are normalized by the maximum light color excess.  Clearly the reddening decreases with time as the scattered photons from the light curve peak add to the SN flux.   While we consider here only a thin shell, the width of the shell will also have an effect \citep{Amanullah_Goobar_2011}.  The general trends of \citet{Wang_2005} and \citet{Goobar_2008}, namely a smaller value of R$_V$ are seen in some of the models with a smaller radius where the scattering delay is neglible.  However, it is clear that a single extinction law cannot be considered representative of circumstellar interaction as it varies with distance, density, and the intrinsic spectral evolution as well as phase.  We have only considered a thin shell--other geometries give qualitatively similar results \citep{Amanullah_Goobar_2011} but which may need to be tested for best results.


\section{SN~2014J compared to Circumstellar Scattering Models\label{comparisons}}
\subsection{Comparison Between Expected and Observed Extinction Evolution}
 
One prediction of the circumstellar scattering model is a negative temporal change in R$_\lambda$.  This is due to the time delay of the scattering and the rising flux of the SN.  
Thus the early light curve will suffer from subtractive extinction but as time passes there should also be an additive component from the scattered light which grows with time until the light curve peak has been scattered back into the line of sight.  The SN light curve will be broader than it would be if observed in isolation.  When compared to an unreddened SN this will appear as a reduction in the R$_\lambda$ as the SN appears less extinguished with time (see Figure 2 in \citealp{Wang_2005}).  We have already shown in Figure \ref{fig_lightcurves} that applying a uniform reddening law with low R$_V$ to the SN~2011fe template does a good job of matching the SN~2014J.  This includes a stretching out of the light curves consistent with the changing effective wavelengths compared to the unreddened SN~2011fe light curves (as also shown by \citealp{Amanullah_etal_2014}).  

\begin{figure} 
\includegraphics[scale=1]{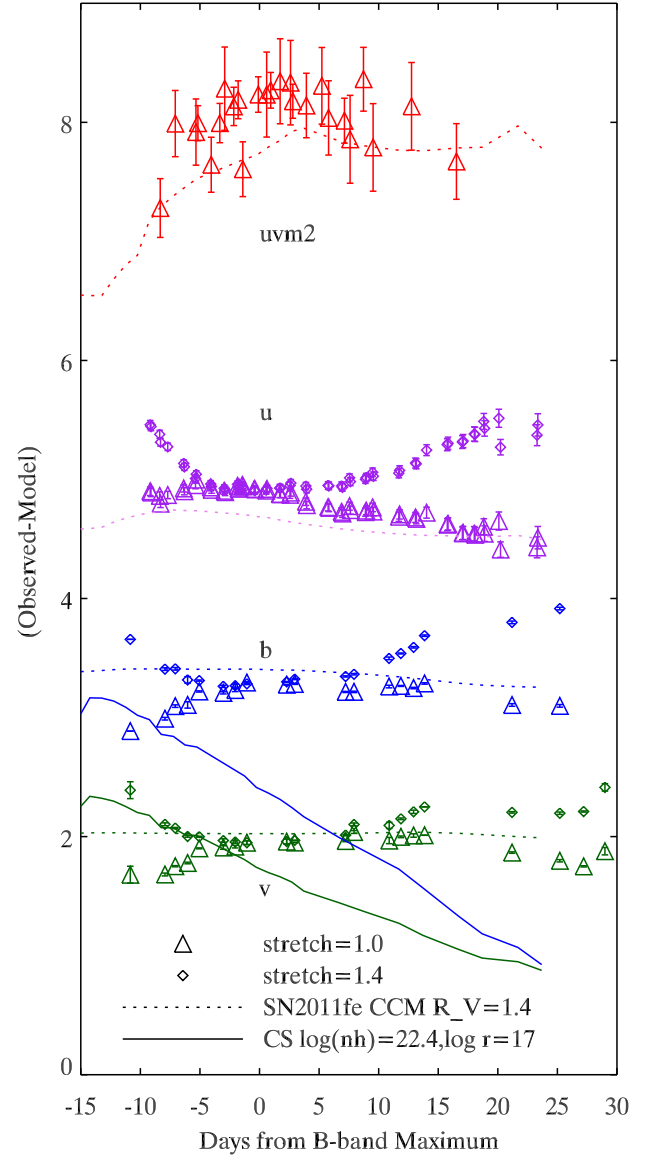}
\caption[Results]
        {
Distance-corrected magnitude differences (effectively the extinction) between SNe 2014J and 2011fe (triangles) and between SN~2014J and a stretched SN~2011fe.  The dotted lines represent the magnitude differences expected from reddening SN~2011fe using the CCM R$_V$=1.4 extinction law with E(B-V)=1.4.  The solid line shows the magnitude difference expected from circumstellar scattering of dust with n$_h$=10$^{22.4}$ and r=10$^{17}$ cm.  
Rather than the monotonic decrease in extinction expected from circumstellar scattering, the SN2014J-SN2011fe differences increase slightly to maximum light.  A better match between the observed magnitude differences and that expected from pure interstellar scattering can be achieved by scaling the epochs of SN~2011fe by a factor of 1.1.  To mimic the early decrease in A$_\lambda$ from scattering one can assume the intrinsic light curve was broader.  After maximum, however, A$_\lambda$ would increase as SN~2014J fades faster than the assumed light curve.   }\label{fig_RU}    
\end{figure}

The UV/optical spectral series of SN~2011fe allows us to directly compare the differences between objects to the magnitude differences caused by extinguishing the evolving spectrum. In Figure \ref{fig_RU} we show with dotted lines the magnitude differences between the reddened and unreddened SN~2011fe spectrophotometry.  The evolution is modest in the optical but is quite significant in the near-UV.  The plotted symbols show the difference between SN~2014J and the SN~2011fe spectrophotometry interpolated to the same epochs.  The difference shows a modestly greater slope than expected from pure spectral evolution.  The difference peaks near maximum light and is smaller at early and late times.  It does not exhibit the rapid fading predicted by the circumstellar scattering \citep{Wang_2005} but is characteristic of SN~2014J having slightly broader light curves than SN~2011fe (corresponding to a relative stretch factor of 1.1).  

To get the temporal decrease in R$_\lambda$  expected from circumstellar scattering, one must assume an intrinsically broader SN as the template.  We simulate this in Figure \ref{fig_RU} by stretching the SN~2011fe template before the interpolation and subtraction.  This results in a premaximum decrease in R$_\lambda$ but also a post-maximum increase in R$_\lambda$ as the post-maximum light curve shape should be stretched out much more than SN~2014J.  

Interestingly, \citet{Foley_etal_2014} do find a temporal change in R$_V$ when comparing UV/optical spectra of SN~2014J to SNe 2011fe and 2013dy.  They cite \citet{Patat_2005} in claiming this as evidence of circumstellar scattering.  But the temporal change found by \citet{Foley_etal_2014} is a positive increase in R$_V$, while \citet{Patat_2005} and \citet{Wang_2005} predict a decrease in R$_V$.  The temporal difference in extinction curves, which increases in the blue (Figure 12; \citealp{Foley_etal_2014}), is also cited in support of the circumstellar scattering.  However, the predictions of \citet{Patat_etal_2006} show such a difference between 6000 and 4000 \AA (Figure 14), while \citet{Foley_etal_2014} see the difference only growing shortward of 4000 \AA.  The UV region is known to be a source of increased variation in SNe Ia \citep{Brown_etal_2010,Milne_etal_2013,Foley_Kirshner_2013}.

\subsection{Direct Comparison with Light Curves}

The change in apparent extinction discussed above is really just the effect of scattered light being added to the observed flux and slowing down the post-maximum decline.  This increased flux is called a light echo at later times (see e.g. \citealp{Patat_2005}) but the effect is the same.  The change in light curve shape near maximum light was also studied by \citet{Amanullah_Goobar_2011} for scattering clouds at various radii. The magnitude of the effect is smaller at smaller radii because the light is scattered back into the line of sight with a short delay time.  The effect is smaller at larger radii because the photons have a much larger delay time and are more spread out in time.   Dust corresponding to E(B-V)$\sim$0.4 at radii between 10$^{16}$ and 10$^{19}$ cm is expected to broaden the light curve such that the measured \mb~is increased by up to several tenths of a magnitude \citep{Amanullah_Goobar_2011}.  Their simulations broaden a normal SN with \mb=1.05 to \mb=0.4, broader than ever observed.    To result in the rather normally broad observed \mb=1.1, SN~2014J would have to have been intrinsically narrow, inconsistent with the observed high-velocity features, otherwise spectroscopic similarity to SN~2011fe and SN~2007co \citep{Foley_etal_2014}, and the absolute magnitudes.  Comparing to other Swift SNe or with the SN~2011fe with different stretch values applied, one can obtain magnitude differences which decrease after maximum light but which are accompanied by a rise before maximum light.  One would have to assume that SN~2014J had an intrinsically long rise time and an intrinsically fast decay in order for scattering to result in the light curve observed.

\begin{figure*} 
\plottwo{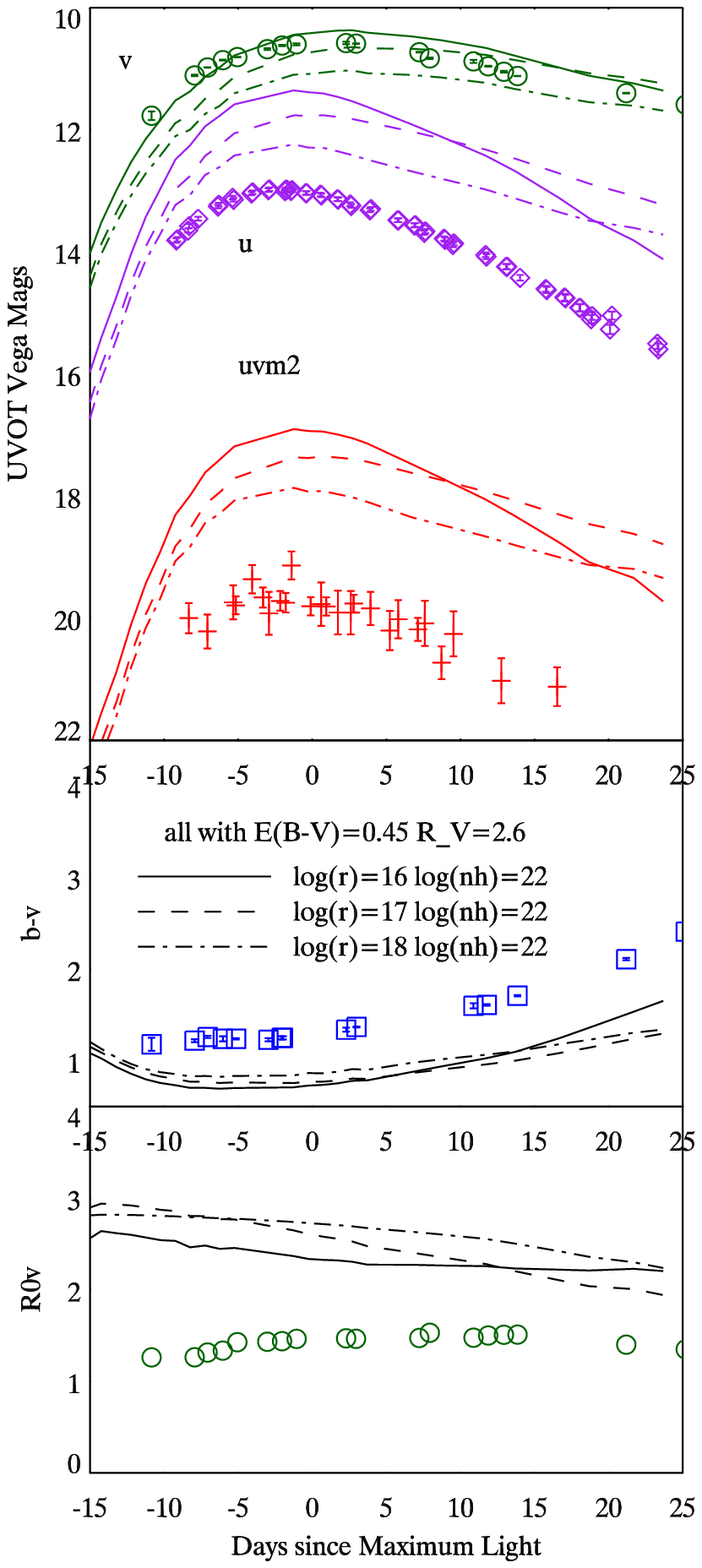}{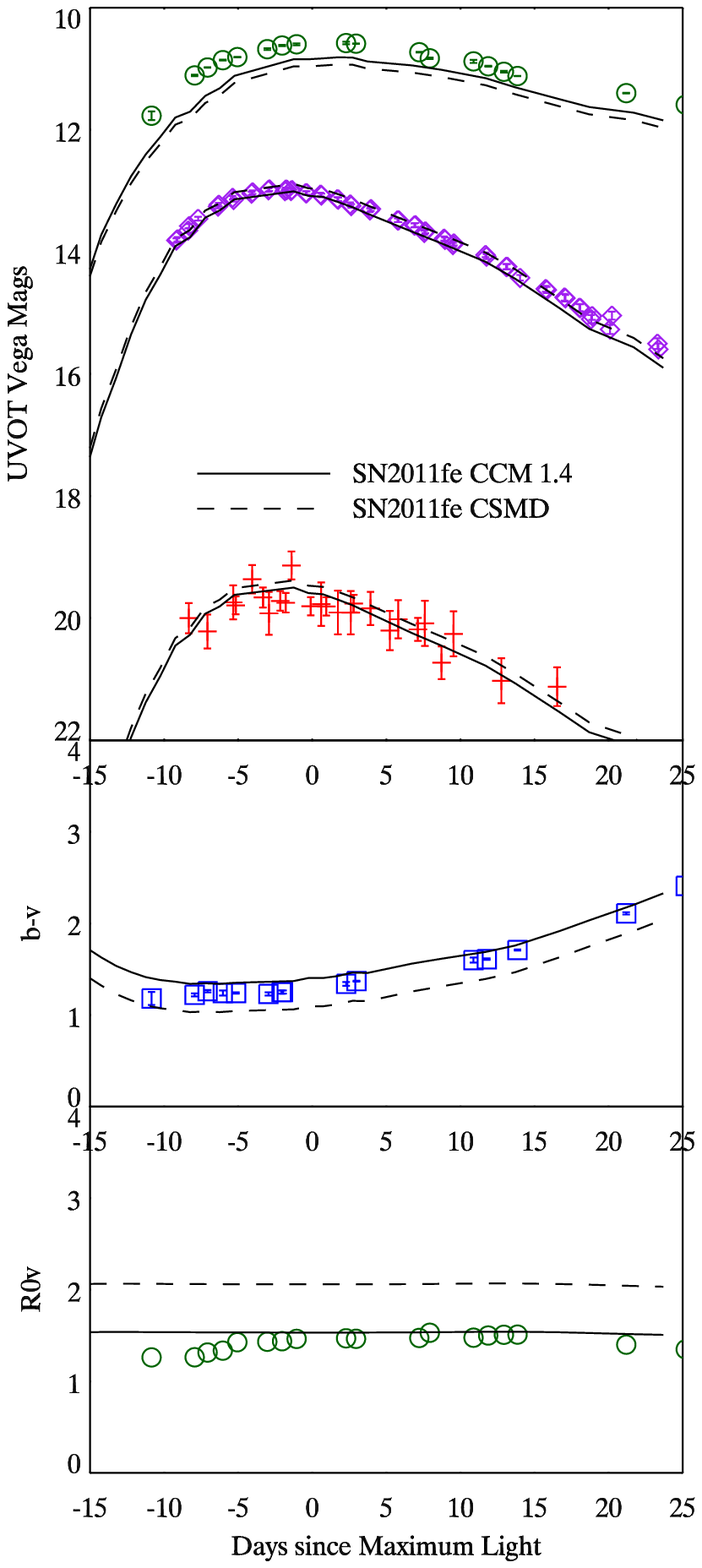}
\caption[Results]
        {Left: The SN2014J photometry, B-V color and R$0_V$ are compared to models with the SN~2011fe series modified by circumstellar dust at different radius values and a foreground reddening of E(B-V)=0.45 and R$_V$=2.6 as suggested by \citet{Foley_etal_2014}.  R$0_V$ is the notation used by \citet{Wang_2005} to designate A$_V$ divided by E(B-V)$_peak$ rather than the color excess at the same epoch.  When approximately matching the V-band magnitude, it is hard to match the other filters, colors, or R$_V$.
Right:  The SN2014J photometry, B-V color and R$0_V$ are compared to a model with SN~2011fe reddened with a Cardelli law with E(B-V)=1.4 and R$_V$=1.4.  We also show the predictions from the \citet{Foley_etal_2014} combination of a Cardelli law with E(B-V)=0.45 and a more reasonable R$_V$=2.6 with a power-law extinction curve similar to \citet{Goobar_2008} with a=0.83, p=-2.6 and E(B-V)=0.6 (called CSMD).

 } \label{fig_csphot}    
\end{figure*} 

We now directly compare our scattering models to the observed light curve.  The scattering model described here is dependent on the hydrogen column density (nh; as discussed above this is used to calculate the optical depth based on the dust model of \citealp{Weingartner_Draine_2001}) and the radius (r) of the dust shell (in our analytic approximation we consider only a thin shell). An additional uncertainty is how much of the reddening is due to circumstellar rather than interstellar dust.  We certainly expect significant interstellar dust in the starburst host M82 \citep{Hutton_etal_2014}. Strong diffuse interstellar bands (DIBs) were detected in SN~2014J spectra \citep{Welty_etal_2014} also suggestive of interstellar dust. 

\citet{Foley_etal_2014} suggested half of the extinction is from interstellar dust.  We ran many simulations of the SN~2011fe spectral series being scattered by many configureations of dust distance and column density.  This was followed by interstellar reddening using the CCM R$_V$=2.6 law with E(B-V)=0.45, as suggested by \citet{Foley_etal_2014}, and the small amount of MW reddening using the CCM R$_V$=3.1 law and E(B-V)=0.054.  In the left panel of Figure \ref{fig_csphot} we show the observed photometry to different models for the scattering light curve.  Of interest is the light curve shapes and the relative offsets between the bands (which represents the colors).  While the light curve shapes can be matched with a smaller radius (10$^{16}$ cm), the colors do not match for any model and R$_V$ is too high.  

As shown above in Figure \ref{fig_RU}, changing the intrinsic colors and light curve shapes and then adding circumstellar scattering could reproduce the observed light curve shapes.  However, the excellent matches of SN~2011fe reddened by a foreground screen of dust, albeit with extreme R$_V$, to the observations of SN~2014J suggest that such fine tuning is not the answer.  In the right panel of Figure \ref{fig_csphot} we show SN~2011fe extinguished by a CCM extinction law with R$_V$=1.4 and E(B-V)=1.4 and the small amount of MW reddening using the CCM R$_V$=3.1 law and E(B-V)=0.054.  We also show a combination of CCM R$_V$=2.6 law with E(B-V)=0.45 and a power-law extinction curve similar to \citet{Goobar_2008} with a=0.83, p=-2.6 and E(B-V)=0.6 (and the MW reddening) suggested by \citet{Foley_etal_2014}.  Both give a reasonable match.   The R$_V$ for the \citet{Foley_etal_2014} combination model is higher, but that is a derived parameter from A$_V$ and B-V, neither of which are significantly worse than the CCM 1.4 model, B-V is simply lower instead of higher.  Agreement is not perfect in the wide range of observational data \citep{Amanullah_etal_2014,Foley_etal_2014}, but there is an intrinsic dispersion in the colors of SNe Ia which could affect our inferred reddening values.  We may also be trying to force an extinction law into particular functional forms which it may not follow.  As discussed above, the fact that some or all of the extinction law can be fit with a power law does not mean it is circumstellar in origin.  But the fitting of the extinction curve with a component of dust with a larger R$_V$ more similar to the LMC or the MW does reduce the amount of dust which would need to be peculiar.  It is likely the case that we do not understand the dust in external galaxies as well as we think we do.

\begin{figure*} 
\plottwo{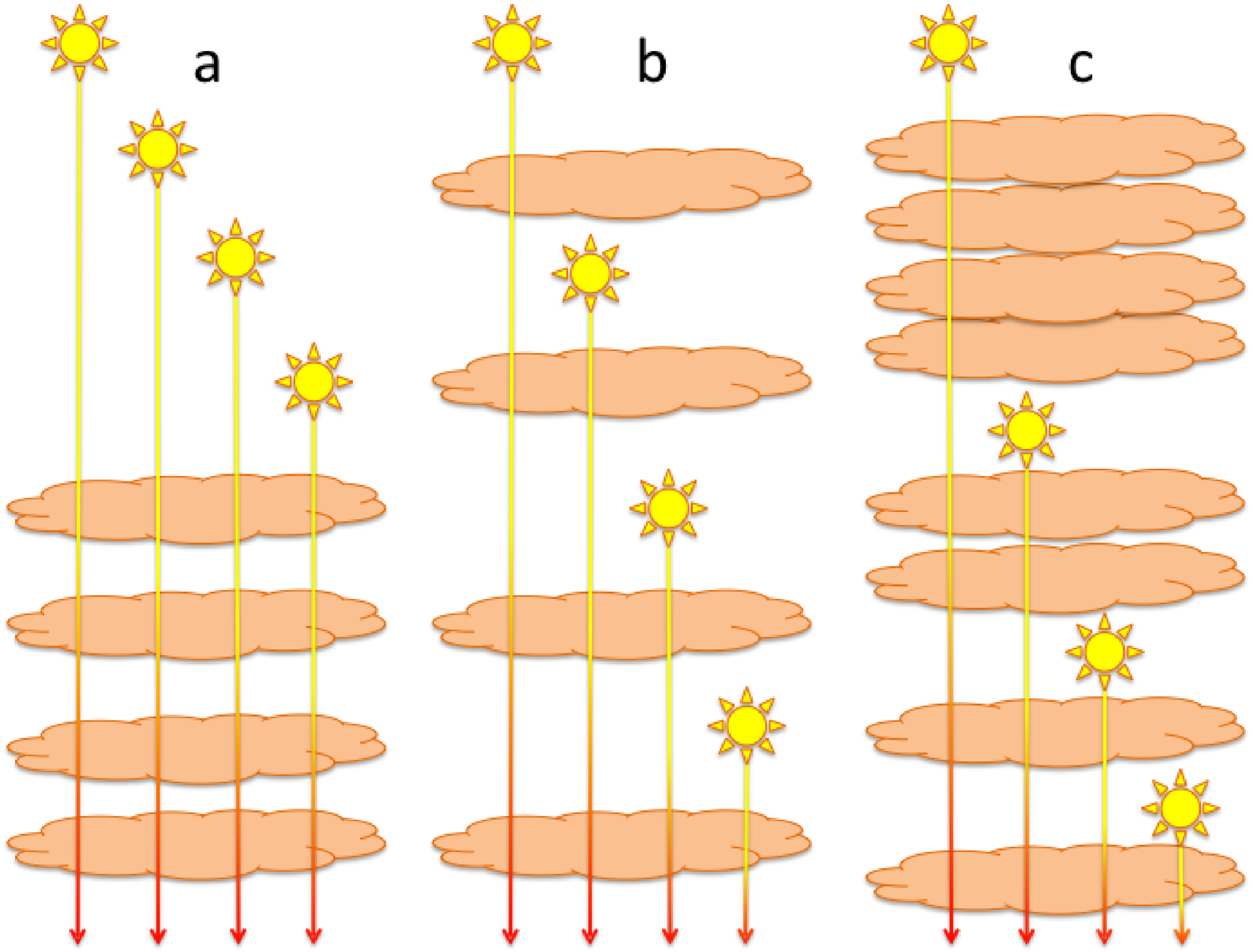}{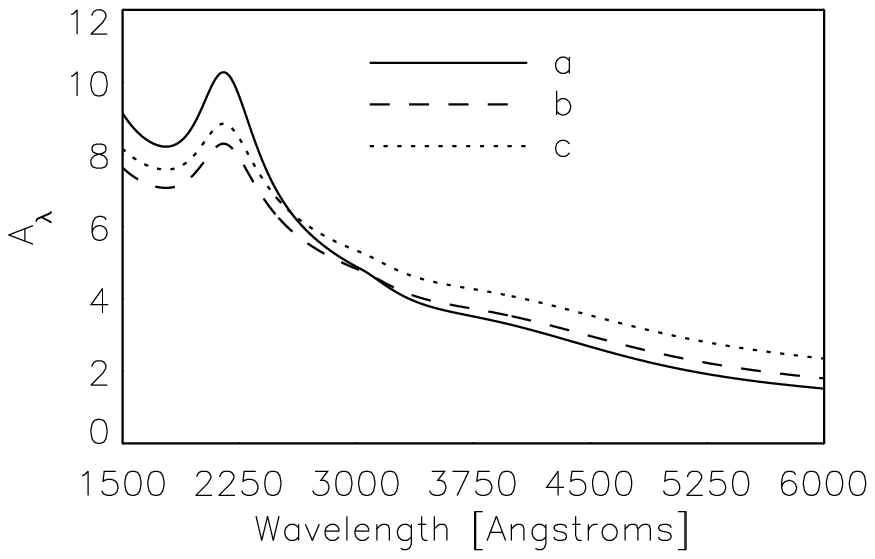}
\caption[Results]
        {
Left:  Different distributions of dust clouds with respect to the stars.  Most extragalactic extinction determinations are calculated as a foreground sheet in front of the stellar content.  Other configureations with the dust mixed within the stellar content, resulting in different amounts of reddening to different stellar populations is more likely.

Right: Inferred reddening laws from the three scenarios presented at the left.  The resulting R$_V$ values are 1.82, 2.12, 2.69.  Mixing the stars and dust results in an increase in the effective R$_V$.

 } \label{fig_clouds}    
\end{figure*}

\section{Galaxy Extinction}\label{galaxyextinction}

If the dust extinction to SN~2014J is interstellar, rather than circumstellar, than why does it seem so different than MW dust?  Much of the effort of explaining the extinction laws toward SNe Ia has assumed it was the SN extinction which is peculiar, requiring differences in the dust geometry or the intrinsic colors of SNe Ia.  However, one could also examine the methods by which galaxy reddening laws are determined.  Maybe a large fraction of external galaxies have ``peculiar'' dust.  

While lines of sight in the MW, LMC, and SMC can be probed by comparing single stars to low reddening counterparts, the extinction laws in other galaxies are usually inferred from the integrated light of stellar populations.  As in the reddening study of \citet{Hutton_etal_2014} for M82, the dust correction is often done assuming a foreground sheet of dust.  Such an assumption is depicted in the left panel of Figure \ref{fig_clouds}.  In reality, the dust is likely mixed in more continuously with the stars.  To test the effect such a change in the geometry would have on the inferred dust extinction law, we show two additional scenarios.  In the first case, four stars (using Vega as the spectral template) are extinguished by four clouds of dust with a combined E(B-V)=1 in the foreground.  The extinction is computed using a Cardelli law with R$_V$=1.7.  A value of R$_V$=1.8 would be measured, not too different from the input value.  The second scenario mixes the stars and dust evenly, such that one star is extinguished by one cloud, one star by two, one star by three clouds, and one star by four clouds.  The effective extinction law is determined by comparing the total transmitted flux to the unextinguished flux of the four stars.  The effective extinction has R$_V$= 2.1.  The third scenario has the dust content increasing more rapidly than the stellar content such that one star is extinguished by one cloud, one by two, one by four clouds, and one by eight clouds.  Assuming that the four stars are extinguished only be foreground dust, the effective extinction has R$_V$=2.7.  So mixing the dust in with the stars increases the effective R$_V$ above what one would measure if it was truly just a foreground dust screen.
This difference in shape is due to differences in the total extinction as well as the difference in extinction between the B and V bands which are used to normalize the curve.  The contribution of scattering was ignored, even along our one dimensional line of sight.  The real situation is even more complicated.

This test was done with a known spectrum for the intrinsic light.  In practice, the stellar mass, effective age or star formation history, and extinction are all being solved for simultaneously.  Even knowing the intrinsic stellar content, assuming a wrong geometry results in an incorrect extinction law.  When the intrinsic stellar content is unknown, then assuming that wrong geometry is unlikely to result in correct stellar masses, ages, and extinction in much more complicated situations.  Stellar population synthesis models should explore the variation caused by different but reasonable assumptions on the dust distribution on the inferred parameters.  Physically motivated dust prescriptions, such as that determined by \citet{Witt_etal_1992} or \citet{Charlot_Fall_2000}, could improve our understanding.  They may also reveal that dust in other galaxies is not as similar to MW dust as has been assumed thus far. 

 \citet{Patat_etal_2014} have studied the polarization to several reddened SNe and found that they do not follow the relations determined for stars in the MW.  Studying the extinction and polarization laws in external galaxies using SNe is similar to the star-matching methods used in the MW.  The circumstellar scattering arguments \citep{Wang_2005,Goobar_2008,Foley_etal_2014} have tried to explain the apparently different colors of reddened SNe with normal dust but an unusual geometry .  However, it may be the dust itself which is unusual, at least compared to what we understand.  A more fundamental point, and one addressed much better by \citet{Calzetti_2001}, is that ``dust obscuration of galaxies is conceptually different from the dust extinction of stars'', and the methods used are very different too.  By comparing a single reddened SN with a similar unreddened comparison, we can probe dust extinction in external galaxies in a similar manner to how MW dust extinction laws are probed.  This is especially the case for SN~2014J,  because the multi-epoch, multi-wavelength comparisons can be done with different subsets and assumptions as shown well by \citet{Amanullah_etal_2014} and \citet{Foley_etal_2014}.  Further data on both reddened and unreddened SNe (since we need comparison objects spanning the properties of the reddened SNe) will improve our knowledge of more galaxies beyond our own.

\section{Summary} \label{discussion}


We have shown that the effective reddening law is dependent on the dust geometry, column density and phase.  This corresponds to the change in extinction and light curve predicted by \citet{Wang_2005} and \citet{Amanullah_Goobar_2011}.  Thus the same reddening law is not applicable to all situations.  The reddening law for a specific dust geometry and epoch may be fit with a power law, but simply matching a power law does not mean that the reddening is from circumstellar dust.  A full scattering model should specify the geometry and optical depth.

By comparing the observed NUV-optical spectra of SN~2014J with SNe 2011fe and 2012fr, we spectroscopically confirm a low value of R$_V$, consistent with that found by others \citet{Goobar_etal_2014,Amanullah_etal_2014,Foley_etal_2014,Marion_etal_2014}. The light curve shapes can be well modeled by reddening the spectral series of SN~2011fe, showing no evidence for contribution from scattered light.  By comparison with models for the temporal signatures of circumstellar scattering, we conclude that little or none of the dust extinction comes from circumstellar dust.  
This is consistent with the clean circumstellar environment suggested by the radio \citep{Perez-Torres_etal_2014} and X-ray \citep{Margutti_etal_2014J} limits and the lack of variation in the narrow sodium absorption lines \citep{Foley_etal_2014}.  While circumstellar dust could still modify the extinction to some SNe Ia, it does not appear to be the cause of the low R$_V$ for SN~2014J.  We have also shown that the assumption of dust extinction as a foreground screen can result in the mischaracterization of the extinction law and likely the inferred stellar population properties.

The high reddening of SN~2014J allowed detailed studies of the near-UV, optical, and near-infrared extinction which were less affected by the intrinsic differences between SNe.  However, it also complicated mid-UV photometry that could otherwise probe the region around the 2175 \AA~bump in many extinction curves.  Probing shorter wavelengths will require mid-UV photometry of less reddened SNe where the effective wavelength is not so dramatically shifted and spectroscopy as distance and sensitivity allow.  The intrinsic differences will also have to be more carefully accounted for, as the variation in SNe Ia increases dramatically in the mid-UV \citep{Brown_etal_2010,Milne_etal_2013,Foley_Kirshner_2013}.
More spectral sequences of SNe Ia in the UV are needed to disentangle the effects of intrinsic and extinction differences in understanding SN colors.



P.J.B. is supported by the Mitchell Postdoctoral Fellowship and 
NASA ADAP grant NNX13AF35G.
L.W. acknowledges support from NSF grant AST-0708873.  
P.A.M. acknowledges support from NASA ADAP grant NNX10AD58G.
A.B., M.P., and N.P.M.K. acknowledge the support of the UK Space Agency.  
This work made use of public data in the {\it Swift} data
archive and the NASA/IPAC Extragalactic Database (NED), which is
operated by the Jet Propulsion Laboratory, California Institute of
Technology, under contract with NASA.

\appendix

\section{Filter Complications} \label{filterconcerns}

\begin{figure} 
\includegraphics[scale=0.6]{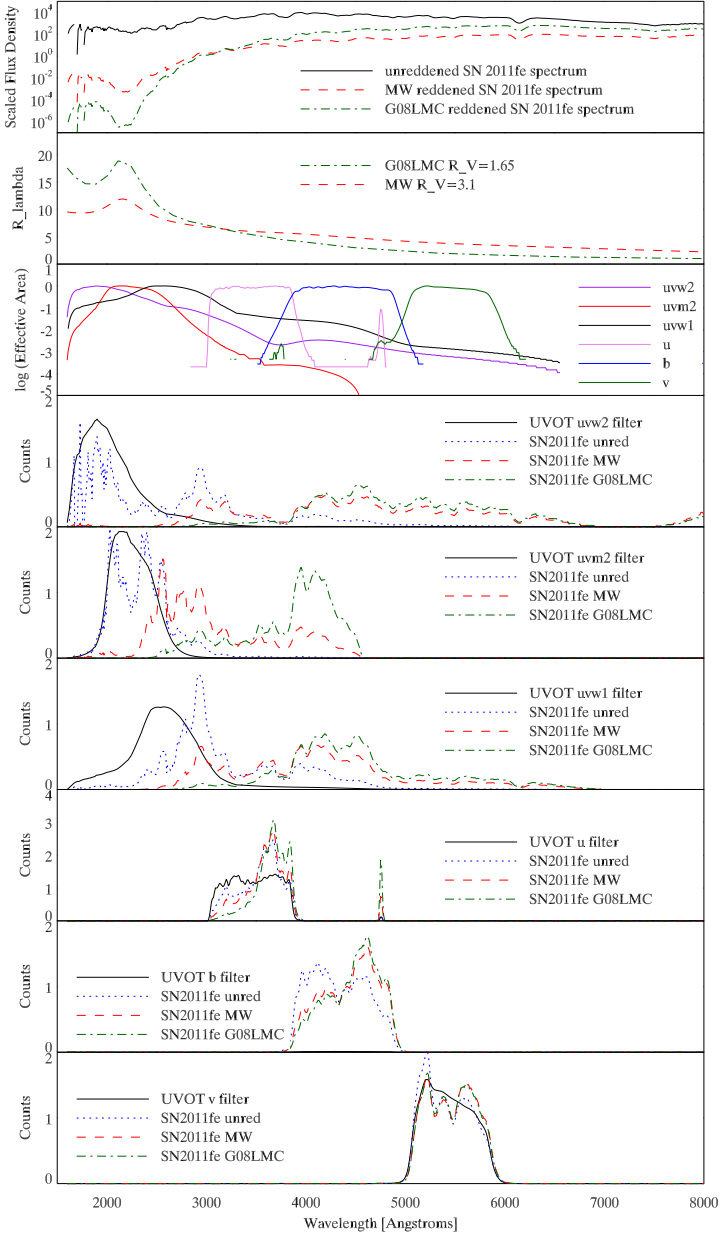} 
\caption[Results]
        { Top Panel: HST UV spectrum of SN2011fe near maximum light: unreddened, and with a Milky Way (MW: R$_V$=3.1) and an LMC extinction law with circumstellar scattering  (G08LMC)\citep{Goobar_2008,Brown_etal_2010} for an E(B-V)=1.2.  
	Second Panel: Wavelength dependence of the MW and G08LMC extinction coefficient.  	Third Panel: UVOT filter effective area curves.
	Bottom Panels: Count spectra after passing the unreddened and reddened spectra through the UVOT filter curves and normalized by the total count rate. 

 }\label{fig_redfilterplots}    
\end{figure}

To properly interpret photometric data, including light curves, colors, and the extinction through a given filter, it is important to understand what is being measured in a photometric data point.  In a count-based detector, all incoming photons resulting in a detected count are treated the same, regardless of wavelength/energy.  So the spectral shape of the source (including reddening) and the total filter+telescope+atmosphere transmission (as appropriate for the instrument) are equally important.  The effect of the optical tails of the uvw2 and uvw1 filters is already recognized \citep{Poole_etal_2008}, especially for the very red spectrum of a SNe Ia \citep{Brown_etal_2010}.  The severe reddening to SN~2014J exacerbates the problems and makes interpretation particularly complicated.  To illustrate the effects, in Figure \ref{fig_redfilterplots} we use the HST STIS spectrum of SN~2011fe from 2011 Sep 13 (the epoch with the broadest wavelength coverage; \citealp{Mazzali_etal_2014}) and apply a reddening of E(B-V) using the MW 3.1 and G08LMC extinction laws, both with an E(B-V)=1.2.  Also shown are the UVOT filter curves and the resulting count spectra when the unreddened and reddened SN spectra are passed through the UVOT filter curves.  The count spectra are normalized by the total number of counts for that spectrum/filter combination (and then scaled uniformly for presentation purposes) to highlight the distribution of the photons with wavelength.

\begin{figure*} 
\plotone{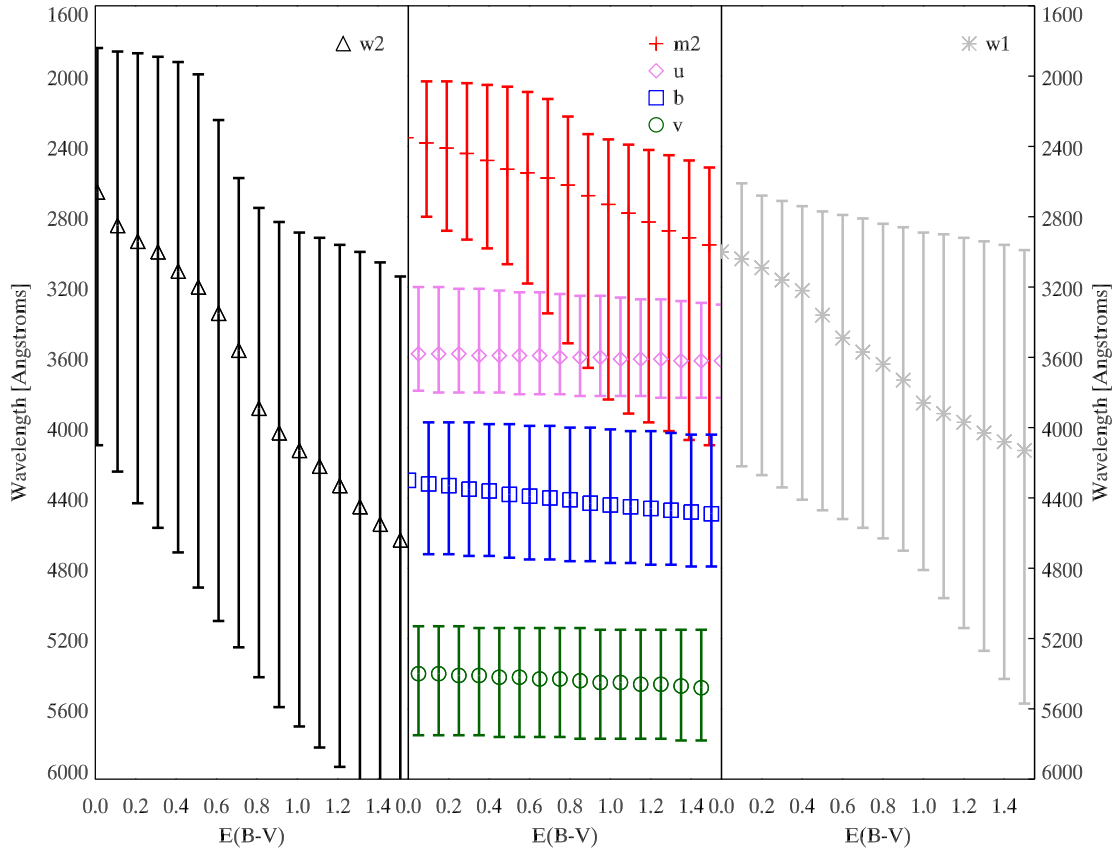} 
\caption[Results]
        {Wavelength ranges of the photons transmitted through the UVOT filters for the SN~2011fe spectrum as a function of reddening (using a MW extinction law with R$_V$=3.1).  The symbols represent the effective wavelength of the filter for the reddened spectrum, defined as the wavelength at which half of the detected photons are received from shorter and longer wavelengths.  The vertical error bars representing the wavelengths between 10 and 90\% of the transmitted photons (i.e., the region within those error bars contain 80\% of the transmitted photons).
 } \label{fig_effwavelengths}    
\end{figure*}

The reddening of the spectrum serves to shift the distribution of observed photons to longer wavelengths.  The effect is small in the optical bands but increases as the wavelength range of the filter decreases.  This general effect is due to the steepening of the extinction laws at shorter wavelengths.  For the UVOT UV filters, the non-negligible tails of the filters dramatically increase the magnitude of the effect.  At a wavelength of 3500 \AA~the uvw2 and uvw1 filters still have about 10\% and 1\% of their peak transmission.  This is already important for an unreddened SN Ia, as they typically have one hundred times more flux at 4000 \AA~ than they do at 2000 \AA.   While the intrinsically very red SN Ia spectrum already has significant counts transmitted through the optical tails of the uvw2 and uvw1 filters, the G08LMC law in particular completely suppresses the remaining mid-UV and near-UV flux, resulting in only optical (B-band) counts being transmitted.
The uvm2 filter has a better cutoff, with only 0.1\% of the peak transmission at 3500 \AA, but, as shown in Figure \ref{fig_redfilterplots}, this could still be significant if the observed source is red enough.  

To further quantify this, Figure \ref{fig_effwavelengths} plots the effective wavelength (essentially the median point for the detected photons for a given filter/spectrum combination) for the Swift UVOT filters for the SN~2011fe spectrum with various amounts of reddening applied.  The wavelengths bounding the central 80\% of the photons are represented by the vertical error bars.  The wavelength ranges of the observed photons for the optical filters are rather constant with reddening due to their sharp cutoffs in transmission.  The UV filters behave very differently.  The optical tails of the uvw2 and uvw1 filters cause the detected photons to have a large range in wavelength which continues to shift to longer wavelengths with increased reddening.
The effective wavelengths for uvw2 and uvw1 match the u band for a reddening of E(B-V)=0.7 and continue to grow.   
Because of the above issues, our analysis above excluded the uvw2 and uvw1 filters and focused on the other UVOT filters: uvm2, u, b, and v.

The uvm2 filter does not have the optical tails of the other UV filters nor does it cut off quite as sharply as the optical filters.  The effective wavelength begins at 2350 \AA~ for an unreddened SN Ia and increases slowly to approach 3000 \AA~ for E(B-V)=1.2.  So while its effective transmission shifts to longer wavelengths into a broader near-UV filter, more than half the photons are still expected to be from wavelengths shorter than the atmospheric cutoff (for a MW 3.1 extinction law).
Clearly the extinction in a given UV band cannot be accurately represented by a magnitude difference at the effective wavelength, as is often done in the optical and NIR (e.g. Figure 14 in \citealp{Folatelli_etal_2010}, Figure 11 in \citealp{Wang_etal_2008}).  
To properly interpret the data, one must compare to spectral templates which have been appropriately reddened.  Spectrophotometry of such models will include the effects of the filters shapes (assuming the filter curves are accurately determined) and allow a fair comparison.


\bibliographystyle{apj}

%

\begin{deluxetable}{llrrrrrrr} 
\tablecaption{UVOT Photometry \label{table_photometry}} 
\tablehead{\colhead{SN} & \colhead{Filter} & \colhead{MJD} & \colhead{Mag} & \colhead{M\_Err} & \colhead{MagLimit} & \colhead{SatLimit} &  \colhead{Rate} & \colhead{R\_Err}   }  
\startdata 

SN2014J & UVW2    &   56679.4377  &  17.050  &   0.088  &  20.095  &  11.091  &   1.355   &  0.110  \\
SN2014J & UVM2    &   56679.4496  &    NULL  &    NULL  &  20.113  &  10.557  &   0.025   &  0.015  \\
SN2014J & UVW1   &    56679.4617  &  15.771  &   0.068  &  19.554  &  11.164  &   4.650  &   0.292  \\
SN2014J & U     &     56680.5633  &  13.814  &   0.039  &  17.873  &  11.061  &  64.624  &   2.323  \\
\enddata 
\tablecomments{  The full table is available in the electronic version.\\
  The photometry will also be available from the Swift SN website \\
http://swift.gsfc.nasa.gov/docs/swift/sne/swift\_sn.html.  } 
\end{deluxetable}

\begin{deluxetable}{lccc}
\tablecaption{Light Curve Parameters\label{table_fits}}
\tablehead{ \colhead{Filter} & \colhead{Peak Magnitude} & \colhead{Peak Date} & \colhead{Delta M 15} \\ 
\colhead{} & \colhead{(mag)} & \colhead{(MJD)} & \colhead{(mag)} } 

\startdata
 \\
uvw2 & 15.800 $\pm$ 0.048 &  56688.9 $\pm$ 1.1 &  0.98 $\pm$ 0.11 \\
uvm2 & 19.54 $\pm$ 0.18 &  56687.9 $\pm$ 0.5 &  0.95 $\pm$ 0.21 \\
uvw1 & 14.594 $\pm$ 0.023 &  56688.0 $\pm$ 0.4 &  0.88 $\pm$ 0.07 \\
u & 12.983 $\pm$ 0.014 &  56687.4 $\pm$ 0.2 &  1.14 $\pm$ 0.03 \\
       \\
\enddata
\tablecomments{b and v fit parameters excluded due to the data being corrupted by high galaxy count rates.}


\end{deluxetable}

\begin{deluxetable}{llll}
\tablecaption{Swift/UVOT Grism Exposures for SNe 2014J and 2012fr\label{table_grism}}
\tablehead{\colhead{SN Phase} & \colhead{Observation ID+extensions} & \colhead{Date} & \colhead{Exposure} \\ 
\colhead{(Days)} & \colhead{ } & \colhead{(UT)} & \colhead{(seconds)} } 
\startdata
SN2014J  -1 & 33123025+2-6 & 2014-01-31 -02-01 & 7518 \\
SN2014J   0 & 33123032+2-4, 33123033+1-4 & 2014-02-02 -3 & 8030 \\
SN2014J  +2 & 33123036+2, 33123037+2 & 2014-02-04  & 4330 \\
SN2014J  +3 & 33123038+1-2, 33123040+1-2  & 2014-02-05  & 4730 \\
SN2012fr -1 & 32614021+1-3, & 2012-11-11 & 4545 \\
SN2012fr +1 & 32614025+1-12, & 2012-11-13  &  16425 \\
\enddata
\end{deluxetable}

\end{document}